\documentclass[conference]{IEEEtran}
\IEEEoverridecommandlockouts
\usepackage{bm}
\usepackage{cite}
\usepackage{amsmath,amssymb,amsfonts}
\usepackage{algorithmic}
\usepackage{graphicx}
\usepackage{textcomp}
\usepackage{epsfig}
\usepackage{times}
\usepackage{longtable}
\usepackage{supertabular,booktabs}
\usepackage{ltablex}
\usepackage{array}
\usepackage{setspace}
\usepackage{xcolor,soul}
\usepackage{balance}
\usepackage{etoolbox}
\usepackage[labelsep=period,font={small}]{caption}

\newcolumntype{C}[1]{>{\centering\arraybackslash} m{#1}}

\def\BibTeX{{\rm B\kern-.05em{\sc i\kern-.025em b}\kern-.08em
    T\kern-.1667em\lower.7ex\hbox{E}\kern-.125emX}}

\newcommand{\qed}{\hfill $\square$}

\renewcommand{\thetable}{\arabic{table}}

\begin{document}

\title{Tracking Error Based Fault Tolerant Scheme for Marine Vehicles with Thruster Redundancy\\
\thanks{This work was partially supported by the Korea Institute of Marine Science \& Technology Promotion (KIMST) funded by the Ministry of Ocean and Fisheries (RS-2024-00432366), also in part by the same organization with the grant No. RS-2023-00256122, all in the Republic of Korea. \\
* Corresponding author: J. H. Li is with the Korea Institute of Robotics and Technology Convergence (KIRO); Tel. +82-54-279-0467.}
}

\author{
\IEEEauthorblockN{Ji-Hong Li*}
\IEEEauthorblockA{\textit{Autonomous Systems R\&D Division} \\
\textit{KIRO}\\
Pohang, Republic of Korea \\
jhli5@kiro.re.kr}
\and
\IEEEauthorblockN{Hyungjoo Kang}
\IEEEauthorblockA{\textit{Autonomous Systems R\&D Division} \\
\textit{KIRO}\\
Pohang, Republic of Korea \\
hjkang@kiro.re.kr}
\and
\IEEEauthorblockN{Min-Gyu Kim}
\IEEEauthorblockA{\textit{Autonomous Systems R\&D Division} \\
\textit{KIRO}\\
Pohang, Republic of Korea \\
mcklee@kiro.re.kr}
\and
\IEEEauthorblockN{Mun-Jik Lee}
\IEEEauthorblockA{\textit{Autonomous Systems R\&D Division} \\
\textit{KIRO}\\
Pohang, Republic of Korea \\
zxdwa0817@kiro.re.kr}
\and
\IEEEauthorblockN{Han-Sol Jin}
\IEEEauthorblockA{\textit{Autonomous Systems R\&D Division} \\
\textit{KIRO}\\
Pohang, Republic of Korea \\
hsjin@kiro.re.kr}
\and
\IEEEauthorblockN{Gun Rae Cho}
\IEEEauthorblockA{\textit{Autonomous Systems R\&D Division} \\
\textit{KIRO}\\
Pohang, Republic of Korea \\
sandman@kiro.re.kr}
}

\maketitle

\begin{abstract}
This paper proposes an active model-based fault and failure tolerant control scheme for a class of marine vehicles with thruster redundancy. Unlike widely used state and parameter estimation methods, where the estimation errors are utilized to generate residual, in this paper we directly apply the trajectory tracking error terms to construct residual and detect thruster fault and failure in the steady state of the tracking system. As for identification or diagnosis, this paper proposes a novel scheme through a detailed examination of the tracking error trends and the combinations of thruster configurations. Since this fault detection and identification operates within the same closed-loop of the tracking control system, control reconfiguration can be easily achieved by adjusting the weight parameter of the isolated thruster to minimize tracking errors or residual. Numerical studies with the real world vehicle model is also carried out to verify the effectiveness of the proposed method.

\end{abstract}


\section{Introduction}
Underwater vehicles have become the most powerful and efficient tools for humans to explore and develop the deep sea environments over the past few decades. Among them, excluding long-range autonomous underwater vehicles (AUVs), most vehicles including ROVs (remotely operated vehicles) and manned submersibles are designed to have redundant thruster configurations so as to increase reliability and security for these expensive underwater robotic systems. And how to design a control system to increase system reliability, especially in the case of thruster faults and failures, has become one of interesting research topics since then.

FDI methods based on analytical redundancy have been widely studied in various complex dynamics systems \cite{b1,b2}. For some of large scale systems where the dynamic modeling is too time-consuming, model-free method, especially using the expert system and artificial neural network, is suitable for their fault-tolerant control \cite{b3,b2}. However, this method usually requires a suitable training set data, which is difficult to be acquired in many of real dynamics systems. The model-based method leverages the idea of generating residuals that reflect the inconsistency between actual and estimated behavior \cite{b2,b4}. State-estimation and parameter-estimation methods are two of most common schemes in this group \cite{b3,b5,b6}. These methods apply various system identification techniques to estimate system states or model parameters and using corresponding estimation errors to construct proper residuals to detect and diagnosis the faults and failures.

This paper focuses on the model-based active fault tolerant control problem for a class of underwater vehicles with thruster redundancy. Quite a number of related works have been done thus far \cite{b7}--\hspace{1sp}\cite{b12}. In \cite{b8}, the state estimation error with all states measurable is used to detect the fault, and further an unknown input-observer is proposed to isolate this fault. In \cite{b9}--\hspace{1sp}\cite{b11}, a sliding-mode observer is used to estimate the unmeasurable states and use the deviation of sliding surface of this observer in the steady state to detect and isolate the thruster fault and failure. As for control allocation or reconfiguration, all the weight parameters for each thruster are tuned until the minimization of the residual is ensured.

For most of the practical underwater vehicles, they are usually equipped with various navigation systems combining with acoustic positioning system, DVL (Doppler velocity log), and IMU (inertial measurement unit). Therefore, from the control engineering point of view, there is no need for additional observer to estimate the vehicle's states. On the other hand, in the case of redundant systems, we can conveniently design a suitable controller to guarantee the various performances in practice \cite{b13,b14}. Building on these two perspectives, this paper directly applies the tracking or regulation errors to construct a residual for detecting thruster faults and failures. For the next stage of identification, we propose a novel diagnosis scheme based on a detailed analysis of tracking error trends in response to variations in each thrust force. A key advantage of this diagnosis method is its ability to easily and quickly identify the faulty thruster by examining the changing trends of given error combinations, such as 2D position and azimuth errors in the horizontal plane. As for control reconfiguration, it is straightforward that the weight parameter corresponding to the identified thruster is adjusted until the residual falls back below a given criteria.

The remainder of the paper is organized as follows. Section 2 presents the underwater vehicle model, thruster fault and failure model, and some other preliminaries. A general backstepping controller is designed in Section 3, while Section 4 describes the detailed concept and components of the proposed FDI algorithm. Section 5 provides numerical studies to demonstrate the effectiveness of proposed scheme, and Section 6 concludes with a brief summary and discusses some of future work.

\section{Problem Statement}
\subsection{Underwater Vehicle Model}
For the convenience of discussion, in this paper we only the vehicle's 3DOF motion on the horizontal plane, whose kinematics and dynamics can be expressed as following \cite{b15}
\begin{align}
\dot{\bm{\eta}}&=\bm{J\nu}, \label{eq1} \\
\dot{\bm{\nu}}&=\bm{F_V}+\bm{B\tau}, \label{eq2}
\end{align}
where $\bm{\eta}=[x,y,\psi]^T$ denotes the vehicle configuration on the horizontal plane in the navigation frame; $\bm{\nu}=[u,v,r]^T$ is the velocity vector in the vehicle's body-fixed frame; $\bm{F_V}\in \Re^{3\times1}$ indicates the vehicle's modeled nonlinear dynamics including hydrodynamic damping, inertia (including added mass), Coriolis and centripetal, and gravitational terms in each of surge, sway, and yaw directions, and is of class $C^2$; control gain matrix $\bm{B}$ is strictly positive definite; $\bm{\tau}=[\tau_u,\tau_v,\tau_r]^T$ is the thrust force and moment vector; $\bm{J}$ denotes the coordinate transformation matrix from body-fixed frame to navigation frame as following
\begin{equation*}
\bm{J}=\begin{bmatrix}\cos\psi&-\sin\psi&0\\ \sin\psi&\cos\psi&0\\0&0&1\end{bmatrix}.
\end{equation*}

\emph{Remark 1}. For the convenience of discussion, this paper does not consider the uncertainty terms (including the cable and sea current dragging components) in the vehicles dynamics (\ref{eq2}). Indeed, in the case of uncertainty terms, both of matched or unmatched ones, it is still easy for us to provide various robust adaptive control schemes to guarantee the system's UUB (uniform ultimate boundedness) stability (\cite{b14,b16} and references therein).

Unlike previous related works \cite{b9}--\hspace{1sp}\cite{b11}, where the control laws are derived in the navigation frame, in this paper we directly solve the trajectory control problem in the vehicle's body-fixed frame as in (\ref{eq2}).

\begin{figure}[!t]
\centerline{\includegraphics[width=5.1cm]{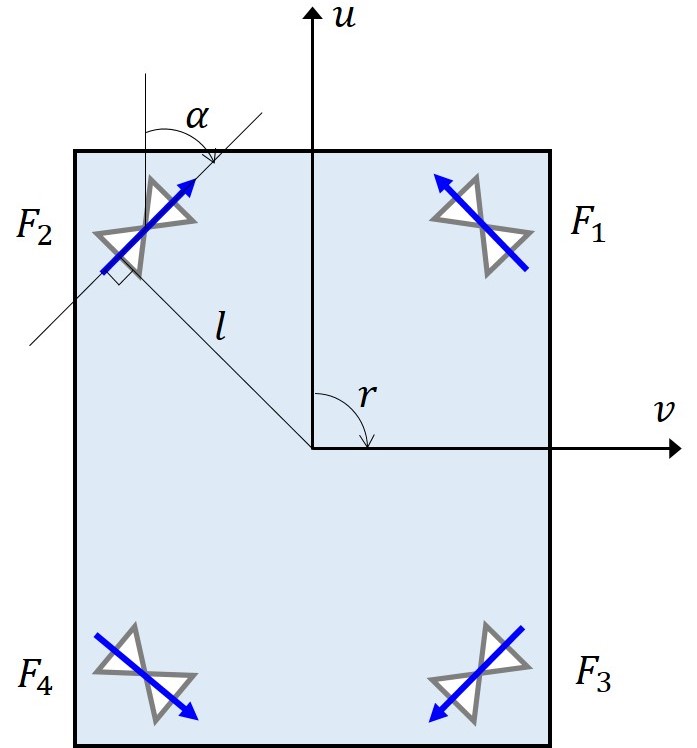}}
\caption{Vehicle thruster configuration.}
\label{fig1}
\end{figure}

As for thruster configuration, we consider the case as in Fig. \ref{fig1}, which is the most common configuration for underwater vehicles. In this case, the control input $\bm{\tau}=[\tau_u,\tau_v,\tau_r]^T$ can be expressed as following
\begin{align}
\bm{\tau}&=\bm{T}_{conf}\bm{F} \nonumber \\
&=\begin{bmatrix}\cos\alpha&\cos\alpha&-\cos\alpha &-\cos\alpha\\ -\sin\alpha&\sin\alpha&-\sin\alpha&\sin\alpha\\-l&l&l&-l\end{bmatrix}\begin{bmatrix}F_1\\F_2\\F_3\\F_4\end{bmatrix}, \label{eq3}
\end{align}
where $\alpha$ is the thruster orientation in the body-fixed frame and $l$ denotes the distance from center point to the thruster orientation line.

\subsection{Thruster Fault and Failure Model}
For each thruster, the output thrust depends on its specific thruster model,
\begin{equation}
F_i=F_T(u_i), \label{eq4}
\end{equation}
where $u_i$ is the thruster actual control input signal. Indeed, this is the input signal into the thruster control board. In the case of BlueROV2 Heavy ROV \cite{b17}, whose model will be used in the numerical studies in this paper, the thruster model (\ref{eq4}) can be embodied as following
\begin{equation}
F_i=K_iu_i, \label{eq5}
\end{equation}
with $K_i$ the thrust coefficient. Here sign of $u_i$ indicates the thrust force direction: If $u_i>0$, output forward thrust; otherwise, output reverse thrust.

As for thruster fault and failure model, this paper directly implements the same model used in \cite{b10,b11} as following
\begin{equation}
F_i=K_iW_iu_i, \label{eq6}
\end{equation}
where $W_i$ is a numerical weight indicating the thrust loss and defined as follows,
\begin{equation}
W_i=\left\{\begin{matrix}0&\!~~~~~\text{if $i$-th thruster has a failure}\\0<W_i<1\!&\!\text{if $i$-th thruster is faulty}\\1&~\text{if $i$-th thruster is normal}\end{matrix}\right. \label{eq7}
\end{equation}

Consequently, the thruster force vector $F$ can be expressed as following
\begin{equation}
\bm{F}=\bm{KWu}, \label{eq8}
\end{equation}
where $\bm{K}=diag(K_1,\cdots,K_4)$, $\bm{W}=diag(W_1,\cdots,W_4)$, and $\bm{u}=[u_1,\cdots,u_4]^T$.

Suppose $\tau_c$ is the control input calculated by the controller, then control allocation \cite{b18} is to allocate the actual control input $\bm{u}$ as following
\begin{equation}
\bm{u}=\bm{W}^{-1}\bm{K}^{-1}\bm{T}_{conf}^{\dag}\bm{\tau}_c, \label{eq9}
\end{equation}
where superscript $``\dag"$ denotes the pseudo-inverse of a non-square matrix.

The problem here is that, in practice, the exact value of $W$ is unknown, and therefore the control allocation has to take the form of
\begin{equation}
\hat{\bm{u}}=\hat{\bm{W}}^{-1}\bm{K}^{-1}\bm{T}_{conf}^{\dag}\bm{\tau}_c, \label{eq10}
\end{equation}
where $\hat{\bm{W}}$ is the estimation of $\bm{W}$ and $\hat{\bm{u}}$ is the corresponding control allocation.

Therefore, the embodied control input $\bm{\tau}$ becomes
\begin{equation}
\bm{\tau}=\bm{T}_{conf}\bm{KW}\hat{\bm{W}}^{-1}\bm{K}^{-1}\bm{T}_{conf}^{\dag}\bm{\tau}_c. \label{eq11}
\end{equation}

If $\hat{\bm{W}}=\bm{W}$, then we have $\bm{\tau}=\bm{\tau_c}$. Otherwise, the deviation of $\hat{\bm{W}}$ from $\bm{W}$ could cause nonzero $\Delta\bm{\tau}=\bm{\tau}_c-\bm{\tau}$, which further result in a similar deviation in the tracking or regulation errors during the steady state of the control system.

\subsection{Problem Formulation}
This paper aims to answer the following question,

\noindent \emph{Is it possible to detect the change in $\bm{W}$ and further identify which $W_i$ is changed, by observing the deviation of the tracking or regulation errors during the steady state of the control system?}

To address this problem, in this paper, we suppose the following constraint conditions.

\emph{Assumption 1}. Thruster fault and failure occur only after the control system has been stabilized.

\emph{Assumption 2}. Thruster fault and failure is caused only by the reduction of corresponding numerical scale weight $\bm{W}$.

\emph{Assumption 3}. At each time, only one thruster occurs fault or failure.

\section{Controller Design}
Suppose $\bm{\eta_d}$ is the vehicle's reference trajectory and corresponding tracking error is defined as $\bm{e}_{\eta}=||\bm{\eta}_d-\bm{\eta}||_2$. Here, if the reference trajectory is taken such that $\dot{\eta}_d=0$, then tracking problem becomes regulation problem. The control objective in this paper is to design a control law for $\bm{\tau}$ such that $\bm{e}_{\eta}(t)\rightarrow0$ with $t\rightarrow\infty$. For this purpose, the vehicle's kinematics (\ref{eq1}) can be rewritten as the following tracking error form
\begin{equation}
\dot{\bm{e}}_{\eta}=\bm{\eta}_d-\bm{J\nu}. \label{eq12}
\end{equation}

For the second-order strict-feedback form of tracking system (\ref{eq12}) and (\ref{eq2}), it is convenient to solve the control problem using general backstepping method \cite{b19}.

\subsection{Kinematics Tracking}
In this step, the following Lyapunov function candidate is considered,
\begin{equation}
\bm{V}_1=\dfrac{1}{2}\bm{e}_{\eta}^T\bm{\Gamma}_1 \bm{e}_{\eta}, \label{eq13}
\end{equation}
where $\Gamma_1>0$ is a diagonal weighting matrix.

By differentiating (\ref{eq13}) and substituting (\ref{eq12}) into it, we have
\begin{equation}
\dot{\bm{V}}_1=\bm{e}_{\eta}^T\bm{\Gamma}_1\left(\dot{\bm{\eta}}_d-\bm{J\nu}\right), \label{eq14}
\end{equation}
according to which, the following control law is chosen for $\bm{\alpha}_{\nu}$, which is the stabilization function \cite{b19} for virtual control input $\bm{\nu}$ in (\ref{eq14}),
\begin{equation}
\bm{\alpha}_{\nu}=\bm{J}^{-1}\left(\dot{\bm{\eta}}_d+\bm{\Gamma}_1^{-1}\bm{A}_1\bm{e}_{\eta}\right), \label{eq15}
\end{equation}
where $\bm{A}_1>0$ is diagonal control gain matrix.

By substituting (\ref{eq15}) into (\ref{eq14}), we can get
\begin{equation}
\dot{\bm{V}}_1=-\bm{e}_{\eta}^T\bm{A}_1\bm{e}_{\eta}+\bm{e}_{\eta}^T\bm{\Gamma}_1\bm{J}\bm{e}_{\nu}, \label{eq16}
\end{equation}
where $\bm{e}_{\nu}=\bm{\alpha}-\bm{\nu}$.

\subsection{Dynamics Tracking}
Now consider the following Lyapunov function candidate
\begin{equation}
\bm{V}_2=\bm{V}_1+\dfrac{1}{2}\bm{e}_{\nu}^T\bm{\Gamma}_2\bm{e}_{\nu}. \label{eq17}
\end{equation}

By differentiating (\ref{eq17}) and substituting (\ref{eq16}) and (\ref{eq2}) into it, we can get
\begin{equation}
\dot{\bm{V}}_2=-\bm{e}_{\eta}^T\bm{A}_1\bm{e}_{\eta}\!+\!\bm{e}_{\eta}^T\bm{\Gamma}_1\bm{J}\bm{e}_{\nu}\!+\!\bm{e}_{\nu}^T\bm{\Gamma}_2\left(\dot{\bm{\alpha}}_{\nu}\!-\!\bm{F}_V\!-\!\bm{B\tau}\right), \label{eq18}
\end{equation}
from which, the final control law for $\bm{\tau}$ is chosen as
\begin{equation}
\bm{\tau}=\bm{B}^{-1}\left(\dot{\bm{\alpha}}_{\nu}-\bm{F}_V+\bm{A}_2\bm{e}_{\nu}+\bm{\Gamma}_2^{-1}\bm{\Gamma}_1\bm{J}\bm{e}_{\eta}\right), \label{eq19}
\end{equation}
where $\bm{A}_2>0$ is diagonal control gain matrix.

\emph{Theorem 1}. Consider the trajectory tracking problem for (\ref{eq1}) and (\ref{eq2}). If we choose the control law as (\ref{eq19}), then we can guarantee the exponential stability of the closed-loop tracking system.

\emph{Proof}. By substituting (\ref{eq19}) into (\ref{eq18}), we have
\begin{align}
\dot{\bm{V}}_2&=-\bm{e}_{\eta}^T\bm{A}_1\bm{e}_{\eta}-\bm{e}_{\nu}^T\bm{A}_2\bm{e}_{\nu} \nonumber \\
&\leq -\lambda\bm{V}_2, \label{eq20}
\end{align}
where $\lambda$ is the minimum singular value of $\bm{\Gamma}_1^{-1}\bm{A}_1$ and $\bm{\Gamma}_2^{-1}\bm{A}_2$. Since $\lambda>0$, this concludes the proof. \qed

\emph{Remark 2}. From (\ref{eq20}), we can see that for any given $\bm{V}_2(0)$ and $\delta>0$, it is always possible to estimate $t_c>0$ such that $||\bm{e}_{\eta}||_2\leq \delta,~\forall t\geq t_c$.

\section{Fault Detection and Identification}
\subsection{Fault Detection}
As mentioned before, this paper constructs a residual using tracking error $\bm{e}_{\eta}$ to detect thruster fault and failure. To distinguish different physical quantities between position error $x_e$ and $y_e$ and Euler angle error $\psi_e$, the residual is defined as
\begin{equation}
R_{detection}=\sqrt{x_e^2+y_e^2+c_1\psi_e^2}, \label{eq21}
\end{equation}
where $c_1>0$ is a weighting factor. Corresponding detection threshold is set as following
\begin{equation}
\delta_{detection}=c_2+f(\dot{\eta}_d,\ddot{\eta}_d), \label{eq22}
\end{equation}
where $c_2>0$ is a weighting factor, and the purpose of the second term $f(\cdot)$ is to account for the smoothness of the reference trajectory $\bm{\eta}_d$.

\emph{Remark 3}. In practice, the most common way of designing a reference trajectory is the way-point method. In this case, at each way-point, the reference trajectory is not smooth, and this often causes the tracking error $\bm{e}_{\eta}$ to spike at these points. The term $f(\cdot)$ in (\ref{eq22}) is included for the purpose of prevent incorrect detection in such special cases.

Now the detection phase becomes straightforward that
\begin{align*}
\small
&\textbf{if}~~ (R_{detection}> \delta_{detection})~~~\text{bTrig}=true; \\
&\textbf{else}~~~~\text{bTrig}=false;
\end{align*}

\subsection{Fault Identification}
Once the detection phase has detected a fault (bTrig= true), the tracking control system triggers an identification task, see Algorithm 1.

\begin{table}[ht]
\small
\centering
\tabcolsep=-0.0cm
\setlength{\arrayrulewidth}{1.4pt}
\begin{spacing}{1.1}
\begin{tabular}{@{}cl@{}} \hline
&\textbf{Algorithm 1:} Fault Identification( ) \\ \hline
&~~\textbf{Input:} $\dot{\bm{e}}_{\eta},\psi,\bm{u},\hat{\bm{W}}$ \\
1&~~$\hat{\bm{W}}$=diag(1,1,1,1); \\
2&~~\textbf{waiting for} (bTrig=true) \\
3&~~~~~\textbf{while}(bTrig) \\
4&~~~~~~~~\textbf{if} (bFirstCheck==true) \\
5&~~~~~~~~~~~faultNum=$bFaultID(\dot{\bm{e}}_{\eta},\bm{u},\psi)$~~~~~~~~~~~~~~~~~~ \\
6&~~~~~~~~~~~bFirstCheck=false; \\
7&~~~~~~~~\textbf{end if} \\
8&~~~~~~~~\textbf{if} (wTime==$T_s$) \\
9&~~~~~~~~~~~$\hat{W}_{faultNum}~-\!\!=\Delta w$; \\
10&~~~~~~~~\textbf{end if} \\
11&~~~~~~~~wTime~+=$\Delta t$; \\
12&~~~~~\textbf{end while} \\
13&~~\textbf{end waiting for} \\
14&~~\textbf{return} false; \\ \hline
\end{tabular}
\end{spacing}
\end{table}

In Algorithm 1, $T_s$ is the weight update time, which depends on the tracking error's convergence rate; $\Delta t$ is the algorithm sampling time; $\Delta w$ is a design parameter that indicates the weight update resolution. The core part of Algorithm 1 is the function $bFaultID(\cdot)$, which identifies the faulty thruster.

\begin{figure}[!t]
\centerline{\includegraphics[width=5.8cm]{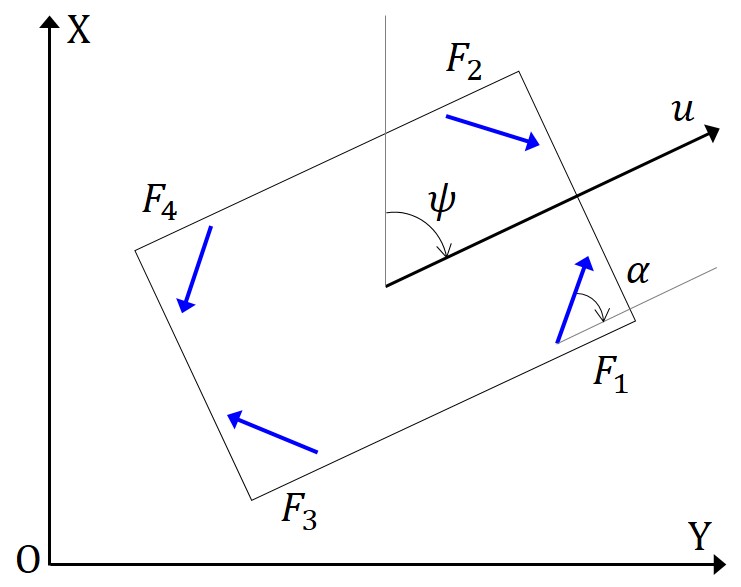}}
\caption{Analysis of the relationship between tracking error change trends and thruster configuration.}
\label{fig2}
\end{figure}

Consider an example case where the vehicle's configuration is shown in Fig. \ref{fig2}. Without loss of generality, here we assume that there is a fault occurred in the first thruster. In this case, all possible cases are analyzed in Table \ref{tab1}, where $\delta_1,\delta_2>0$ are design parameters and depend on the control system performance. Here we take Case 3 in the table as an example for explaining the whole idea in Table \ref{tab1}. In this case: Since $u_1>0$, which means that thrust force is in a forward direction as seen in Fig. \ref{fig2}, reduction of $w_1$ causes a similar reduction in the anti-clockwise yaw moment and this further causes $\dot{\psi}_e$ to be negative; $\cos(\psi-\alpha)<0$ combining with $u_1>0$ indicates the thrust force along X-axis is in the negative direction, therefore the reduction in the thrust force causes $\dot{x}_e$ to be negative; since $\sin(\psi-\alpha)>0$ with $u_1>0$, the thrust force along Y-axis is in the positive way, the reduction in the thrust force causes increasing of $\dot{y}_e$. For any given condition $(\dot{\bm{e}}_{\eta},\bm{u},\psi)$, if there is one case listed in Table \ref{tab1} being satisfied, then we can determine that thruster 1 has occurred fault. And this is the main idea of the function $bFaultID(\cdot)$. This kind of diagnosis applies to all four thrusters at each time in the loop of \emph{Fault Identification( )}, as seen in Algorithm 1.

\begin{table}[ht]
\small
\renewcommand{\arraystretch}{1.2}
\centering
\caption{Case analyses in the case of $W_1$ being reduced.}\label{tab1}
  \tabcolsep=.5cm
\begin{tabular}{@{}C{0.9cm}@{}|@{}C{0.9cm}@{}|@{}C{1.6cm}@{}|@{}C{1.6cm}@{}|@{}C{1.1cm}@{}|@{}C{1.1cm}@{}|@{}C{1.1cm}@{}} \hline\hline
Case & $u_1$ & $\cos(\psi\!-\!\alpha)$ & $\sin(\psi\!-\!\alpha)$ & $\dot{x}_e$ & $\dot{y}_e$ & $\dot{\psi}_e$ \\ \hline\hline
1 & $>0$ & $>0$ & $>0$ & $>\!\delta_1$ & $>\!\delta_1$ & $<\!-\!\delta_2$ \\ \hline
2 & $>0$ & $>0$ & $<0$ & $>\!\delta_1$ & $<\!-\!\delta_1$ & $<\!-\!\delta_2$ \\ \hline
3 & $>0$ & $<0$ & $>0$ & $<\!-\!\delta_1$ & $>\!\delta_1$ & $<\!-\!\delta_2$ \\ \hline
4 & $>0$ & $<0$ & $<0$ & $<\!-\!\delta_1$ & $<\!-\!\delta_1$ & $<\!-\!\delta_2$ \\ \hline
5 & $<0$ & $>0$ & $>0$ & $<\!-\!\delta_1$ & $<\!-\!\delta_1$ & $>\!\delta_2$ \\ \hline
6 & $<0$ & $>0$ & $<0$ & $<\!-\!\delta_1$ & $>\!\delta_1$ & $>\!\delta_2$ \\ \hline
7 & $<0$ & $<0$ & $>0$ & $>\!\delta_1$ & $<\!-\!\delta_1$ & $>\!\delta_2$ \\ \hline
8 & $<0$ & $<0$ & $<0$ & $>\!\delta_1$ & $>\!\delta_1$ & $>\!\delta_2$ \\ \hline\hline
\end{tabular}
\end{table}

\section{Numerical Study}
To verify the effectiveness of the proposed FDI scheme, numerical studies are carried out in this section. In the simulation, as mentioned before, we directly apply the 6DOF nonlinear model of BlueROV2 Heavy ROV \cite{b17}.

\begin{figure}[!t]
\centerline{\includegraphics[width=\columnwidth]{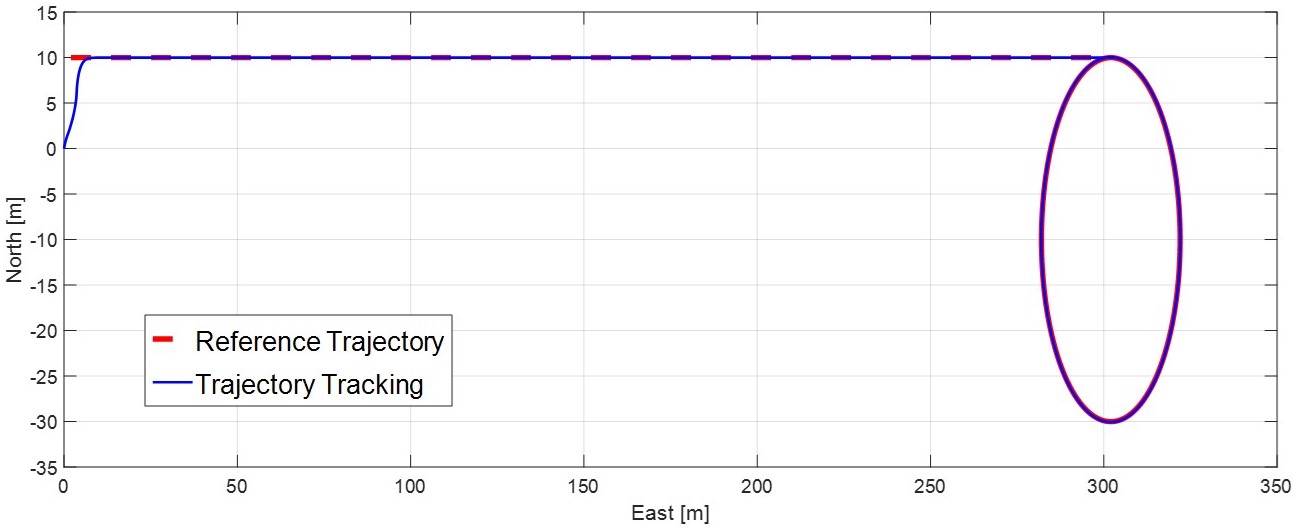}}
\caption{Reference trajectory and its tracking w/o thruster fault.}
\label{fig3}
\end{figure}

\begin{figure}[!t]
\centerline{\includegraphics[width=\columnwidth]{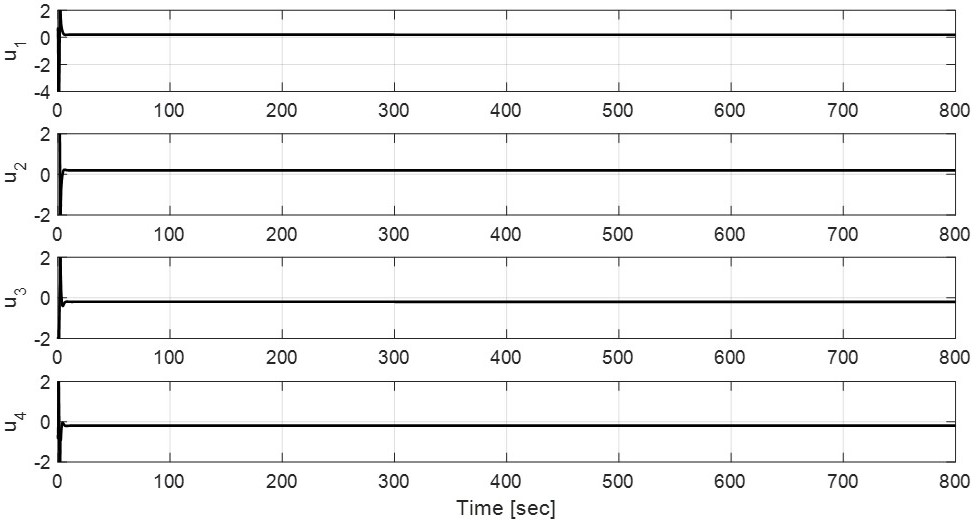}}
\caption{Actual control input $\bm{u}$ w/o thruster fault.}
\label{fig4}
\end{figure}

\begin{figure}[!t]
\centerline{\includegraphics[width=\columnwidth]{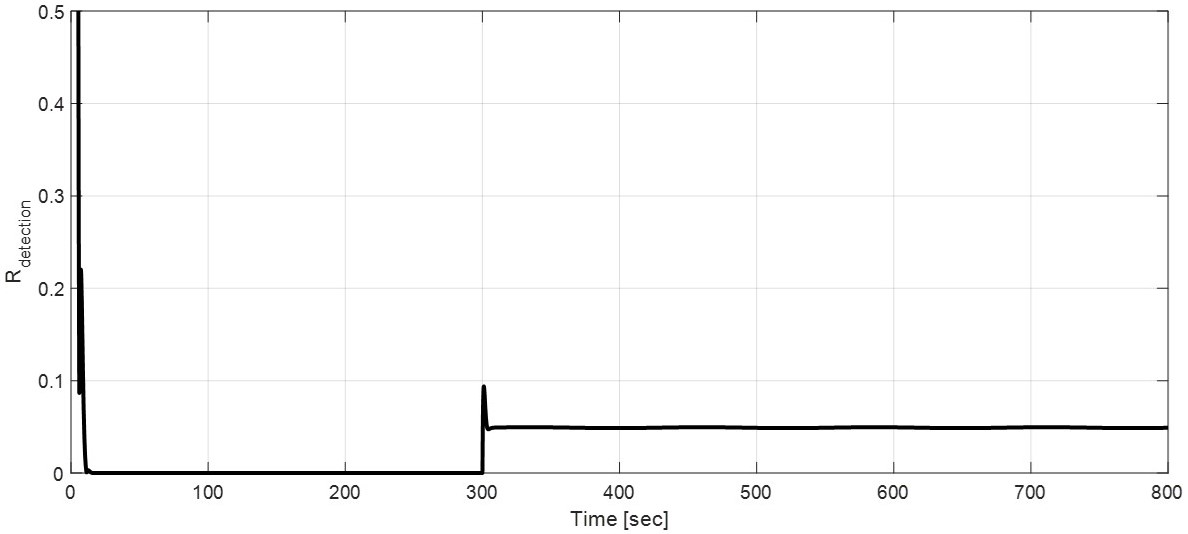}}
\caption{Residual trend in the tracking w/o thruster fault.}
\label{fig5}
\end{figure}

At first, for the vehicle's trajectory tracking controller design, the reference trajectory is set as: at $t<300s$, $\eta_d=[10,5+u_dt,\pi/2]^T$ with $u_d=1.0m/s$; otherwise, $u_d=1.0m/s$ and $r_d=0.05rad/s$. The vehicle's initial state is set as $X=\bm{0}^{12\times1}$, and the controller design parameters are chosen as: $\Gamma_1=diag(1,1,10)$, $\Gamma_2=diag(100,100,300)$, $A_1=diag(1,1,10)$, and $A_2=diag(100,100,300)$. Fig. \ref{fig3} shows the reference trajectory and its tracking with proposed controller in the case of without any thruster fault and failure, and Fig. \ref{fig4} presents the corresponding actual control input $\bm{u}$. Here it is worth to mention that the thrust coefficients are taking $K_i=40$ in (\ref{eq5}) \cite{b17}. The residual defined as (\ref{eq21}) is depicted in Fig. \ref{fig5} with $c_1=5$, from which we can see that the magnitude of the tracking error depends on the smoothness of the trajectory, as described in (\ref{eq22}). In the simulation, we set $c_2=0.01$ and $f(\dot{\bm{\eta}}_d,\ddot{\bm{\eta}}_d)=0.3$. It is worth to mention that, in Fig. \ref{fig5}, there is a residual jumping at $t=300$. This is because at this point, the trajectory is not smooth. This residual jumping is also accounted for designing $\delta_{detection}$ in (\ref{eq22}).

\begin{figure}[!t]
\centerline{\includegraphics[width=\columnwidth]{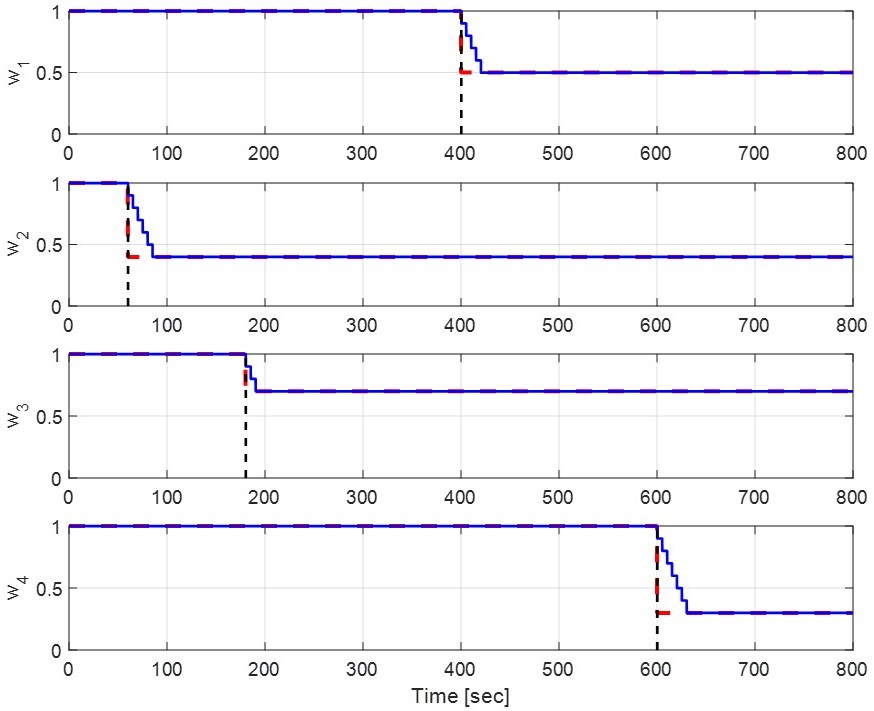}}
\caption{Fault identification. Dotted red line indicates actual fault weights, solid blue line is estimated fault weights, and vertical dotted black line denotes the time of fault detection.}
\label{fig6}
\end{figure}

\begin{figure}[!t]
\centerline{\includegraphics[width=\columnwidth]{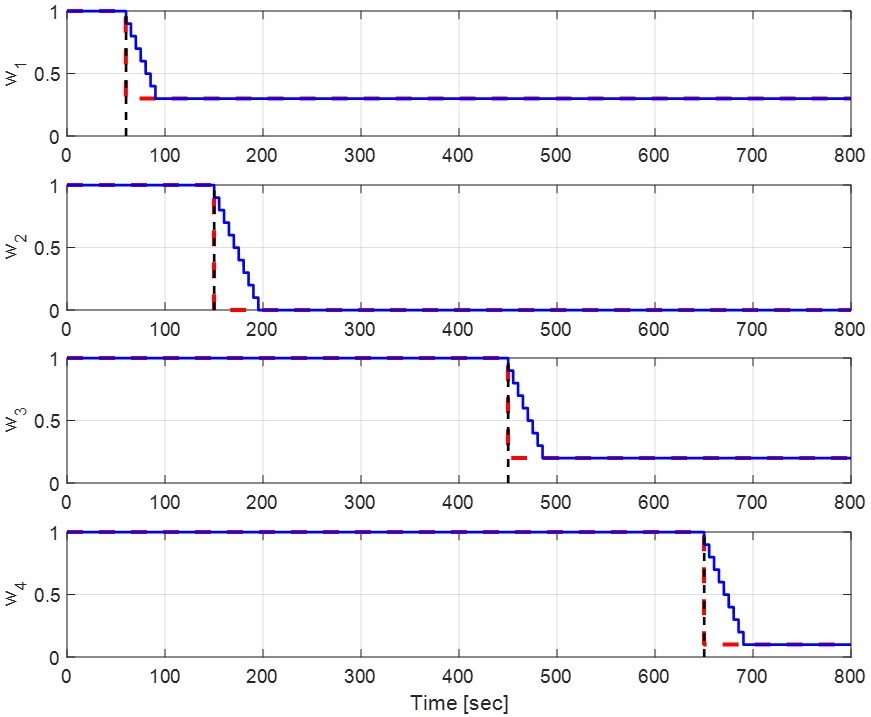}}
\caption{Fault identification with one thruster failure. Dotted red line indicates actual fault weights, solid blue line is estimated fault weights, and vertical dotted black line denotes the time of fault detection.}
\label{fig7}
\end{figure}

\begin{figure}[!t]
\centerline{\includegraphics[width=\columnwidth]{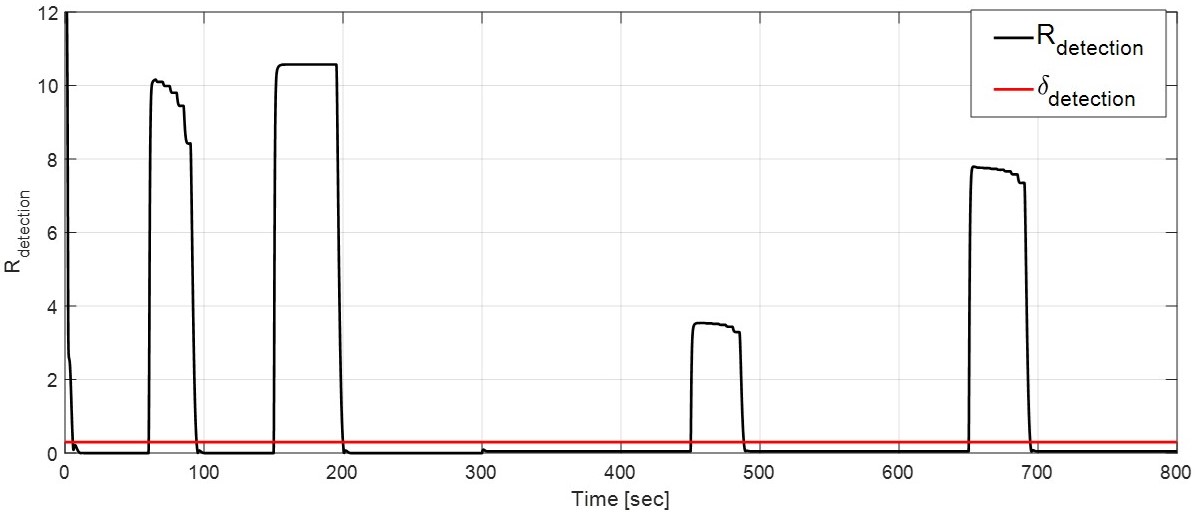}}
\caption{Residual trend in the same case of Fig. \ref{fig7}.}
\label{fig8}
\end{figure}

\begin{figure}[!t]
\centerline{\includegraphics[width=\columnwidth]{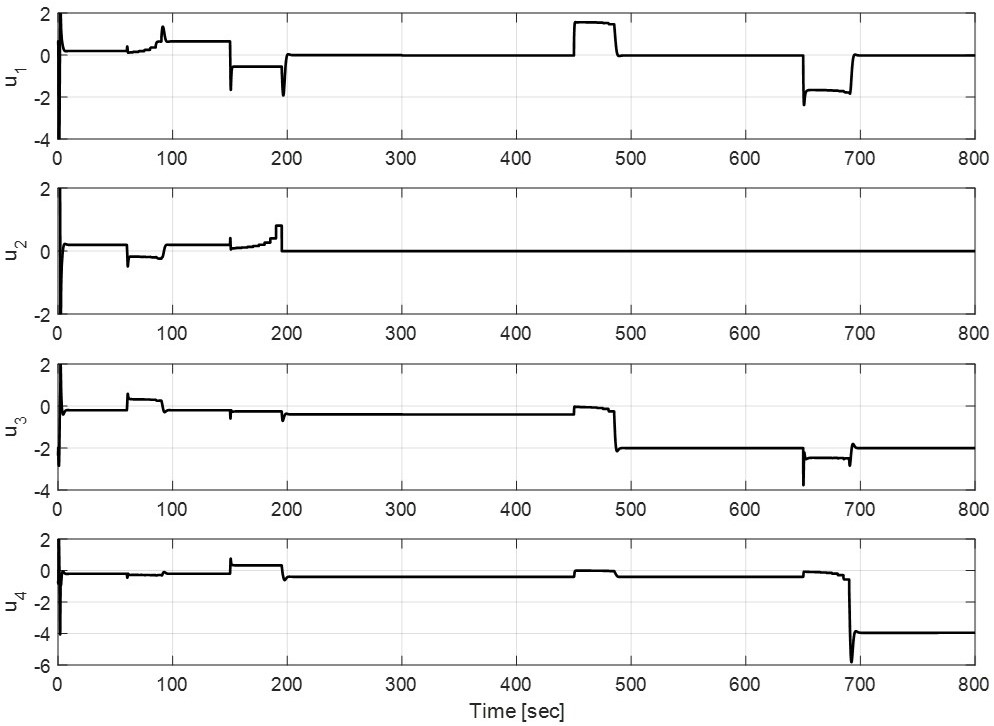}}
\caption{Actual control inputs in the case of Fig. \ref{fig7}.}
\label{fig9}
\end{figure}

To verify the proposed FDI algorithm, various simulations have been carried out, among which Fig. \ref{fig6} shows the result in the case where faults occur sequentially in the four thrusters, and Fig. \ref{fig7} presents the case where there is one thruster failure. From Fig. \ref{fig7}, we can see that in the case of one thruster failure, remained three thrusters are sufficient to guarantee the satisfactory tracking performance even with the small weight values, as $w_1=0.3$, $w_3=0.2$, and $w_4=0.1$. The residual variation and control inputs corresponding to the case Fig. \ref{fig7} are shown in Fig. \ref{fig8} and Fig. \ref{fig9}.

\begin{figure}[!t]
\centerline{\includegraphics[width=\columnwidth]{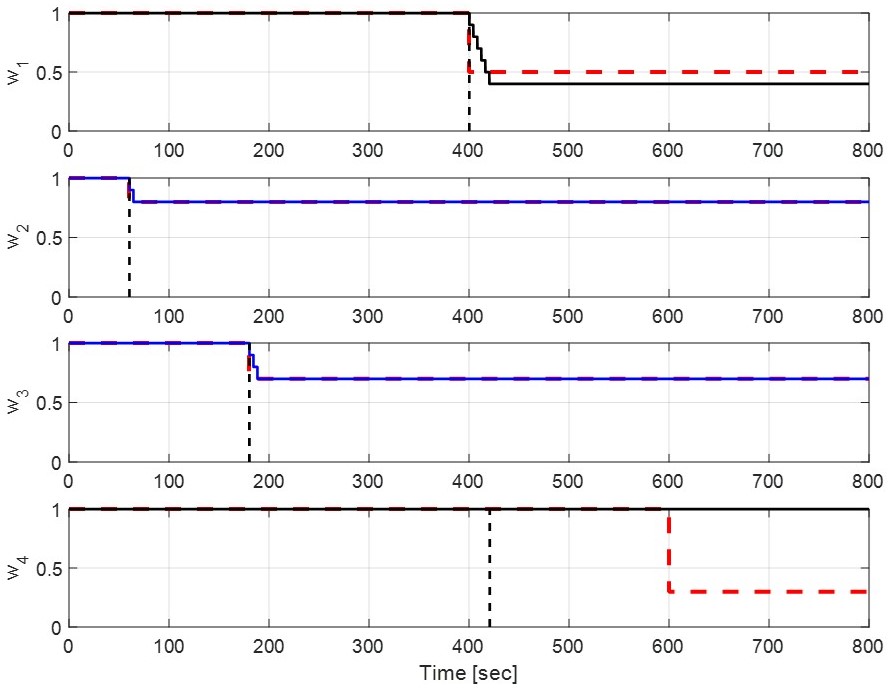}}
\caption{FDI identification failure with small value of weight update time ($T_s=4s$).}
\label{fig10}
\end{figure}

In Algorithm 1, as mentioned before, the parameter $T_s$, which is taken as $T_s=5s$ in the simulation, depends on the control system's tracking performance (or convergence rate). Therefore, if this parameter is set too small, then the algorithm may not work well. And the simulation result shown in Fig. \ref{fig10} just demonstrates this issue. In practice, most work-class ROVs and manned submersibles have relatively slow response speeds, which may necessitate longer $T_s$. Fortunately, this slow response speed usually results in also relatively slow divergence of tracking errors. Therefore, a long $T_s$ might not cause severe problems in most of practical applications.

\section{Conclusion}
This paper has presented an active model-based fault tolerant control scheme for a class of marine vehicles with thruster redundancy. The main idea is directly utilizing the control error to construct a residual and further detect the thruster fault. Through detailed investigations and analyses of the relationship between these control error changing trends and the individual thruster output changing, a novel scheme of fault identification has been proposed. Numerical studies with the real world vehicle model also has been carried out to demonstrate the effectiveness of the proposed scheme.

Despite the satisfactory simulation results, the following issues should be included in our future works:
\begin{itemize}
\item Uncertainty terms, both of matched and unmatched components including sea current and tether cable effects, should be considered and corresponding stable robust adaptive control scheme might be required. In this case, undoubtedly, the proposed FDI method should be revised significantly to meet the requirements.
\item Some restricting conditions, especially \emph{Assumption} 2 and 3, need to be relaxed.
\end{itemize}

\end{document}